\documentclass[aps,article]{revtex4}%
\usepackage{amsfonts}
\usepackage{amsmath}
\usepackage{amssymb}
\usepackage{graphicx}%
\begin{document}

\title{Two photon decay of neutral scalars below 1.5 GeV in a chiral model for
$\overline{q}q$ and $\overline{q}\overline{q}qq$ states}
\author{Sim\'{o}n Rodr\'{\i}guez, Mauro Napsuciale}
\address{Instituto de F\'{\i}sica, Univ. de Guanajuato,
Lomas Del Bosque 103, Fracc. Lomas del Campestre,
37150, Le\'{o}n, Guanajuato, M\'{e}xico. }

\begin{abstract}
We study the two photon decay of neutral scalars below 1.5 GeV in the context
of a recently proposed chiral model for $\overline{q}q$ and $\overline
{q}\overline{q}qq$ states. We find good agreement with experimental results for 
the $a_{0}(980)\rightarrow\gamma\gamma$. Our calculations for  
$f_{0}(980)\rightarrow\gamma\gamma$ shows that further work is necessary in 
order to understand the structure of this meson. The model predicts 
$\Gamma(a_{0}(1450)\rightarrow\gamma\gamma)=0.16\pm 0.10KeV, ~
\Gamma(\sigma   \rightarrow\gamma\gamma)=0.47\pm 0.66\text{ KeV},~
\Gamma(f(1370)   \rightarrow\gamma\gamma)=0.07\pm 0.15\text{ KeV},~
\Gamma(f(1500)    \rightarrow\gamma\gamma)=0.74\pm 0.78\text{ KeV}$.

\end{abstract}

\maketitle

\section{Introduction}

The understanding of the lowest lying scalar mesons remains as one of the
most challenging problems in low energy QCD. Over the past years
experimental evidence has accumulated for the existence of a light scalar
nonet \cite{nonete} but there is still an intense debate about the structure
of these mesons. There are essentially two proposals for the structure of
light scalar mesons: a $q\bar{q}$ structure and a $\bar{q}\bar{q}qq$ one. In
the latter case the are at least three possible dynamical configurations: a
meson-meson molecule, a diquark-diquark state and a compact genuine $\bar{q}%
\bar{q}qq$ state. In addition there exist the possibility that isosinglet
scalars contain a certain amount of glueball. Most of recent work has been
dedicated to the understanding the properties of light scalars in the light
of these pictures, explored in different formalisms.

Since photons couple to charge, electromagnetic decays of mesons have been
intensively used in the past to obtain information on their structure. As
for scalar mesons, there exist calculations for the $a_{0}(980),$ $%
f_{0}(980)\rightarrow 2\gamma $ decays using a variety of approaches \cite%
{est1,est2,Barnes:1985cy,est3}, in particular, in different versions of the
quark model \cite{est1}, with very different results depending on the
details of the model. The generally accepted conclusion, is that the
measured $a_{0}(980),f_{0}(980)\rightarrow \gamma \gamma $ decay widths \cite%
{PDG04} are not consistent with a $q\bar{q}$ structure. In the light of
these results, other possibilities for the structure of light scalars such
as a molecule picture \cite{est2} and a $\bar{q}\bar{q}qq$ structure \cite%
{est3} were explored.

More recently, some tools to determine the glueball content of mesons from
their branching fractions in radiative quarkonium decays and production
cross sections in $\gamma\gamma$ collisions, were developed in \cite{Close}%
. Also, the two photon decay of the lightest scalar, the $f_{0}(600)$ or $%
\sigma$, has been studied using different formalisms \cite%
{Black,Eef02,Rekalo} yielding very disparate predictions. The possibility
that this meson has a large glueball content was analyzed in \cite%
{Pennington} using the two photon decay channel. As shown in this work, the
extraction of the two photon coupling of light isoscalars from data on $%
\gamma\gamma\rightarrow\pi\pi$ is not straightforward and requires a careful
analysis in order to get reliable results.

The two photon decay of the $a_{0}(980)$ and $f_{0}(980)$ mesons were also
studied in \cite{LN} in the framework of a chiral theory involving scalars
and a linear realization of chiral symmetry, a linear sigma model (L$\sigma $%
M), where the key interaction is the one violating $U_{A}(1)$ symmetry \cite%
{NTt}. This interaction is assumed to be the manifestation at the hadron
level of the effective six-quark interaction (for $N_{F}=3$) induced by
instantons\cite{tHooft}. In this concern this interaction identifies the
fields entering the model as $q\bar{q}$ states. In this formalism, the two
photon decay of neutral scalars is induced at the one loop level. The scalar
decay firstly into two (real or virtual) charged mesons which annihilate
into two photons. As to the information on the internal structure of scalars
which can be inferred in this calculation, we must stress that a naive
estimation of the distances explored by the photons in this decay are of the
order of \ $d\approx 1/k$ $=2/m_{S}=0.4\ fm$ which is of the same order as
the kaon charge radius, thus the effective degrees of freedom seen by the
photons are actually mesons rather than quarks. In this sense, we can infer
that the decaying mesons are $q\bar{q}$ states but this bare state has large
quantum fluctuations into meson-meson states whose exact amount is difficult
to quantify.

Recently we pointed out the existence of a quasi-degenerate chiral nonet in
the energy region around $1.4\ GeV$ and studied the possibility that mesons
below 1.5 GeV arise as admixtures of normal $q\bar{q}$ and $\bar{q}\bar{q}qq$
states, with the latter lying at its natural scale as dictated by the linear
rising of meson masses with the number of constituent quarks \cite{NR,NR1}.
This model has the nice feature of reducing to the one explored in \cite{NTt}
in the case when we decouple the $\bar{q}\bar{q}qq$ states.

In this work we explore the predictions of the model presented in \cite%
{NR,NR1} for the two photon decay widths of all neutral scalars below $1.5\
GeV$.

\section{Meson loop contributions to $S\rightarrow\protect\gamma\protect%
\gamma.$}

The most general form for the $S(p)\rightarrow\gamma(k,\epsilon)\ \gamma
(q,\eta)$ transition amplitude is dictated by Lorentz covariance and gauge
invariance as%
\begin{equation}
\mathcal{M}(S\rightarrow\gamma(k,\epsilon)\ \gamma(q,\eta))=\frac{i\alpha}{%
\pi f_{K}}V^{S}\left( g^{\mu\nu}k\cdot q-k^{\mu}q^{\nu}\right) \eta_{\mu
}\epsilon^{\nu},  \label{ampl}
\end{equation}
where $f_{K}$ denotes the kaon weak decay constant and $\alpha$ denotes the
electromagnetic fine constant. The factors $\alpha,\pi,$ $f_{K}$ in Eq.(\ref%
{ampl}) are introduced just for convenience in future manipulations. With
this normalization the form factor $V^{S}(p^{2})$ is dimensionless.

In the effective theory formulated in \cite{NR} the two photon decays of
neutral scalar mesons below 1.5 GeV are induced by loops of charged mesons
as depicted in Fig. 1. A straightforward calculation yields%
\begin{equation}
V_{M}^{S}=\frac{2f_{K}g_{SMM}}{m_{S}^{2}}\left[ -\frac{1}{2}+\xi
_{M}^{S}I(\xi _{M}^{S})\right] ,
\end{equation}%
where $\xi _{M}^{S}=\frac{m_{M}^{2}}{m_{S}^{2}},$ $g_{SMM}$ denotes the
coupling constant of the decaying scalar $S$ to the meson pair ($M^{+}M^{-}$
) in the loops and $I(x)$ denotes the loop integral%
\begin{equation}
I(x)=\left\{ 
\begin{array}{ll}
2\left( \arcsin \sqrt{\frac{1}{4x}}\right) ^{2} & x>\frac{1}{4} \\ 
2\left[ \frac{\pi }{2}+i\ln \left( \sqrt{\frac{1}{4x}}+\sqrt{\frac{1}{4x}-1}%
\right) \right] ^{2} & x<\frac{1}{4}%
\end{array}%
\right. .
\end{equation}%
The decay width is given by%
\begin{equation}
\Gamma (S\rightarrow \gamma \gamma )=\frac{\alpha ^{2}}{64\pi ^{3}}\frac{%
m_{S}^{3}}{f_{K}^{2}}\left\vert V^{S}\right\vert ^{2}.
\end{equation}

\begin{figure}[tbp]
\begin{center}
\includegraphics[
height=2.3281in,
width=3.4921in]{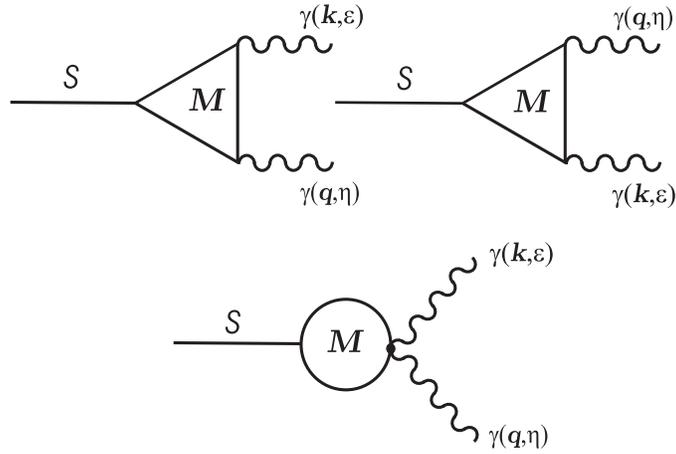}
\end{center}
\caption{Charged meson loop contributions to $S\rightarrow \protect\gamma 
\protect\gamma $}
\label{loop}
\end{figure}
The experimental data for $a_{0}(980),$ $f_{0}(980)$ decay into two photons
is \cite{PDG04} 
\begin{align*}
\Gamma (a_{0}(980)& \rightarrow \gamma \gamma )\times
BR(a_{0}(980)\rightarrow \pi ^{0}\eta )=0.24_{-0.07}^{+0.08}\ KeV \\
\Gamma (f_{0}(980)& \rightarrow \gamma \gamma )=0.39_{-0.13}^{+0.10}\ KeV,
\end{align*}%
and assuming that the decay $a_{0}(980)\rightarrow \pi \eta $ is the dominant
mode ($BR(a_{0}(980)\rightarrow \pi ^{0}\eta )\simeq 1$) we obtain the form
factors $\left\vert V^{a(980)}\right\vert $ and $\left\vert
V^{f(980)}\right\vert $ at $p^{2}=m_{S}^{2}$ as%
\begin{equation*}
\left\vert V_{Exp}^{a(980)}\right\vert =0.34\pm 0.05,\quad \left\vert
V_{Exp}^{f(980)}\right\vert =0.44\pm 0.07.
\end{equation*}%
There exists no confident experimental information for the two photon decays
of the $a_{0}(1450)$, $f_{0}(600)$, $f_{0}(1370)$ and $f_{0}(1500)$ mesons 
\cite{PDG04}.

\section{Two photon decay of $a_{0}(980)$ and $a_{0}(1450)$.}

For the $a_{0}(980)\rightarrow \gamma \gamma $ and $a_{0}(1450)\rightarrow
\gamma \gamma $ decays the model in \cite{NR} yields contributions coming
from are $K$, $\kappa (900)$ and their respective heavy \textquotedblleft
partners\textquotedblright\ $K(1469)$ and $K_{0}^{\ast }(1430)$ in the
loops. For the sake of simplicity we denote these mesons as $K$, $\kappa $, $%
\hat{K}$ and $\hat{\kappa}$ respectively. The couplings entering in 
the loops were calculated in \cite{NR1} and are listed in Table I, we refer
 the reader to \cite{NR1} for details of the notation.

\begin{center}
Table I
\end{center}

\begin{equation*}
\begin{tabular}{|l|l|}
\hline
$g_{a(980)K^{+}K^{-}}$ & $-\frac{1}{\sqrt{2}\left( a+b\right) }\left(
m_{K}^{2}+m_{\hat{K}}^{2}-m_{a}^{2}-m_{A}^{2}+\hat{\mu}_{1}^{2}-\hat{\mu}%
_{1/2}^{2}\right) \cos \phi _{1}\cos ^{2}\theta _{1/2},$ \\ \hline
$g_{a(980)\kappa ^{+}\kappa ^{-}}$ & $\frac{1}{\sqrt{2}\left( b-a\right) }%
\left( m_{\kappa }^{2}+m_{\hat{\kappa}}^{2}-m_{a}^{2}-m_{A}^{2}+\hat{\mu}%
_{1}^{2}-\hat{\mu}_{1/2}^{2}\right) \cos \phi _{1}\cos ^{2}\phi _{1/2},$ \\ 
\hline
$g_{a(1450)K^{+}K^{-}}$ & $-\frac{1}{\sqrt{2}\left( a+b\right) }\left(
m_{K}^{2}+m_{\hat{K}}^{2}-m_{a}^{2}-m_{A}^{2}+\hat{\mu}_{1}^{2}-\hat{\mu}%
_{1/2}^{2}\right) \sin \phi _{1}\cos ^{2}\theta _{1/2},$ \\ \hline
$g_{a(1450)\kappa ^{+}\kappa ^{-}}$ & $\frac{1}{\sqrt{2}\left( b-a\right) }%
\left( m_{\kappa }^{2}+m_{\hat{\kappa}}^{2}-m_{a}^{2}-m_{A}^{2}+\hat{\mu}%
_{1}^{2}-\hat{\mu}_{1/2}^{2}\right) \sin \phi _{1}\cos ^{2}\phi _{1/2},$ \\ 
\hline
$g_{a(980)\hat{K}^{+}\hat{K}^{-}}$ & $-\frac{1}{\sqrt{2}\left( a+b\right) }%
\left( m_{K}^{2}+m_{\hat{K}}^{2}-m_{a}^{2}-m_{A}^{2}+\hat{\mu}_{1}^{2}-\hat{%
\mu}_{1/2}^{2}\right) \cos \phi _{1}\sin ^{2}\theta _{1/2},$ \\ \hline
$g_{a(980)\hat{\kappa}^{+}\hat{\kappa}^{-}}$ & $\frac{1}{\sqrt{2}\left(
b-a\right) }\left( m_{\kappa }^{2}+m_{\hat{\kappa}}^{2}-m_{a}^{2}-m_{A}^{2}+%
\hat{\mu}_{1}^{2}-\hat{\mu}_{1/2}^{2}\right) \cos \phi _{1}\sin ^{2}\phi
_{1/2},$ \\ \hline
$g_{a(1450)\hat{K}^{+}\hat{K}^{-}}$ & $-\frac{1}{\sqrt{2}\left( a+b\right) }%
\left( m_{K}^{2}+m_{\hat{K}}^{2}-m_{a}^{2}-m_{A}^{2}+\hat{\mu}_{1}^{2}-\hat{%
\mu}_{1/2}^{2}\right) \sin \phi _{1}\sin ^{2}\theta _{1/2},$ \\ \hline
$g_{a(1450)\hat{\kappa}^{+}\hat{\kappa}^{-}}$ & $\frac{1}{\sqrt{2}\left(
b-a\right) }\left( m_{\kappa }^{2}+m_{\hat{\kappa}}^{2}-m_{a}^{2}-m_{A}^{2}+%
\hat{\mu}_{1}^{2}-\hat{\mu}_{1/2}^{2}\right) \sin \phi _{1}\sin ^{2}\phi
_{1/2}.$ \\ \hline
\end{tabular}%
\ 
\end{equation*}%
It is worth noticing that the coupling of light scalars to light
pseudoscalars quoted above reduce to the ones obtained in the linear sigma
model \cite{NTt} \ when we decouple the heavy fields. The two photon decays
are of course dominated by light meson in the loops, heavy meson
contributions being suppressed by the corresponding inverse mass powers. The
values extracted in \cite{NR} for the mixing angles entering these
expressions are 
\begin{equation*}
\begin{tabular}{|l|l|}
\hline
Angle & Prediction \\ \hline
$\theta _{1}$ & $18.16^{\circ }\pm 4.3^{\circ }$ \\ \hline
$\theta _{1/2}$ & $22.96^{\circ }\pm 4.8^{\circ }$ \\ \hline
$\phi _{1}$ & $39.8^{\circ }\pm 4.5^{\circ }$ \\ \hline
$\phi _{1/2}$ & $46.7^{\circ }\pm 9.5^{\circ }$ \\ \hline
$\gamma $ & $-9.11^{\circ }\pm 0.5^{\circ }$ \\ \hline
$\delta $ & $21.45^{\circ }\pm 6.5^{\circ }$ \\ \hline
$\delta ^{\prime }$ & $51.36^{\circ }\pm 8.3^{\circ }$ \\ \hline
\end{tabular}%
.
\end{equation*}%
We use also the relations%
\begin{equation}
a=\frac{f_{\pi }}{\sqrt{2}\cos (\theta _{1})},\qquad a+b=\frac{\sqrt{2}f_{K}%
}{\cos (\theta _{1/2})}.
\end{equation}%
Using these values and the masses quoted in \cite{PDG04} we obtain the
results listed in Table II for the form factors. We include the
contributions of different mesons step by step in order to have an idea on
the effects of different mesons in the loops.

\begin{center}
Table II
\end{center}

\begin{equation*}
\begin{tabular}{|l|l|l|l|l|}
\hline
Contr. & $K$ & $K,\kappa $ & $K,\kappa ,\hat{K}$ & $K,\kappa ,\hat{K},\hat{%
\kappa}$ \\ \hline
$\left\vert V^{a(980)}\right\vert $ & $0.331\pm 0.078$ & $0.321\pm 0.080$ & $%
0.323\pm 0.080$ & $0.320\pm 0.084$ \\ \hline
$\left\vert V^{a(1450)}\right\vert $ & $0.149\pm 0.045$ & $0.154\pm 0.051$ & 
$0.153\pm 0.051$ & $0.154\pm 0.053$ \\ \hline
\end{tabular}%
\ 
\end{equation*}%
As expected, the main contribution comes from kaon loops due to its
relatively light mass. These results are in good agreement with the world
average in the case of the two photon decay of the $a_{0}(980).$ The
modifications introduced by the mixing between $\overline{q}q$ and $%
\overline{q}\overline{q}qq$ fields to the picture in the conventional
(updated) linear sigma model \cite{LN} are also exhibited in Table
III.\newpage

\begin{center}
Table III
\end{center}

\begin{equation*}
\begin{tabular}{|l|l|l|l|}
\hline
Form Factor & L$\sigma $M & This model & Exp. \\ \hline
$\left\vert V^{a(980)}\right\vert $ & $0.348\pm 0.038$ & $0.320\pm 0.084$ & $%
0.34\pm 0.05$ \\ \hline
$\left\vert V^{a(1450)}\right\vert $ & \ \ \ \ \ --- & $0.154\pm 0.053$ & \
\ \ \ \ --- \\ \hline
\end{tabular}%
\ 
\end{equation*}%
The form factor for to the $a_{0}(1450)$ decay as predicted by the chiral
model for $\overline{q}q$ and $\overline{q}\overline{q}qq$ states
corresponds to a width 
\begin{equation}
\Gamma (a_{0}(1450)\rightarrow \gamma \gamma )=0.16\pm 0.10\ KeV.
\end{equation}

\section{Two photon decay of isosinglet scalars.}

Next, we work out the predictions of the model for $f_{0}(600)$ (or $\sigma $%
)$,$ $f_{0}(980),$ $f_{0}(1370)$ and $f_{0}(1370)$ decay into two photons.
In this case we expect the main contribution to come from $\pi ,$ $K$ and $%
\kappa $ meson loops. In Table IV we list the couplings involved in these
processes as arising from the chiral model for $\overline{q}q$ and $%
\overline{q}\overline{q}qq$ states

\begin{center}
Table IV
\end{center}

\begin{equation*}
\begin{tabular}{|l|l|}
\hline
$g_{\sigma \pi ^{+}\pi ^{-}}$ & $\frac{1}{\sqrt{2}a}(m_{\sigma }^{2}+m_{\hat{%
\sigma}}^{2}-m_{\pi }^{2}-m_{\hat{\pi}}^{2}+\hat{\mu}_{1}^{2}-\hat{\mu}%
_{1/2}^{2})\cos \gamma \cos \delta \cos ^{2}\theta _{1},$ \\ \hline
$g_{\sigma K^{+}K^{-}}$ & $\frac{1}{\sqrt{2}(a+b)}\left( m_{\sigma }^{2}+m_{%
\hat{\sigma}}^{2}-m_{K}^{2}-m_{\hat{K}}^{2}\right) \left( \cos \gamma -\sqrt{%
2}\sin \gamma \right) \cos \delta \cos ^{2}\theta _{1/2},$ \\ \hline
$g_{\sigma \kappa ^{+}\kappa ^{-}}$ & $-\frac{1}{\sqrt{2}(b-a)}\left(
m_{\sigma }^{2}+m_{\hat{\sigma}}^{2}-m_{\kappa }^{2}-m_{\hat{\kappa}%
}^{2}\right) \left( \cos \gamma +\sqrt{2}\sin \gamma \right) \cos \delta
\cos ^{2}\phi _{1/2},$ \\ \hline
$g_{\sigma \hat{\pi}^{+}\hat{\pi}^{-}}$ & $\frac{1}{\sqrt{2}a}(m_{\sigma
}^{2}+m_{\hat{\sigma}}^{2}-m_{\pi }^{2}-m_{\hat{\pi}}^{2}+\hat{\mu}_{1}^{2}-%
\hat{\mu}_{1/2}^{2})\cos \gamma \cos \delta \sin ^{2}\theta _{1},$ \\ \hline
$g_{\sigma \hat{K}^{+}\hat{K}^{-}}$ & $\frac{1}{\sqrt{2}(a+b)}\left(
m_{\sigma }^{2}+m_{\hat{\sigma}}^{2}-m_{K}^{2}-m_{\hat{K}}^{2}\right) \left(
\cos \gamma -\sqrt{2}\sin \gamma \right) \cos \delta \sin ^{2}\theta _{1/2},$
\\ \hline
$g_{\sigma \hat{\kappa}^{+}\hat{\kappa}^{-}}$ & $-\frac{1}{\sqrt{2}(b-a)}%
\left( m_{\sigma }^{2}+m_{\hat{\sigma}}^{2}-m_{\kappa }^{2}-m_{\hat{\kappa}%
}^{2}\right) \left( \cos \gamma +\sqrt{2}\sin \gamma \right) \cos \delta
\sin ^{2}\phi _{1/2},$ \\ \hline
$g_{f(980)\pi ^{+}\pi ^{-}}$ & $\frac{1}{\sqrt{2}a}(m_{f}^{2}+m_{\hat{f}%
}^{2}-m_{\pi }^{2}-m_{\hat{\pi}}^{2}+\hat{\mu}_{1}^{2}-\hat{\mu}%
_{1/2}^{2})\sin \gamma \cos \delta ^{\prime }\cos ^{2}\theta _{1},$ \\ \hline
$g_{f(980)K^{+}K^{-}}$ & $\frac{1}{\sqrt{2}(a+b)}\left( m_{f}^{2}+m_{\hat{f}%
}^{2}-m_{K}^{2}-m_{\hat{K}}^{2}\right) \left( \sin \gamma +\sqrt{2}\cos
\gamma \right) \cos \delta ^{\prime }\cos ^{2}\theta _{1/2},$ \\ \hline
$g_{f(980)\kappa ^{+}\kappa ^{-}}$ & $-\frac{1}{\sqrt{2}(b-a)}\left(
m_{f}^{2}+m_{\hat{f}}^{2}-m_{\kappa }^{2}-m_{\hat{\kappa}}^{2}\right) \left(
\sin \gamma -\sqrt{2}\cos \gamma \right) \cos \delta ^{\prime }\cos ^{2}\phi
_{1/2},$ \\ \hline
$g_{f(980)\hat{\pi}^{+}\hat{\pi}^{-}}$ & $\frac{1}{\sqrt{2}a}(m_{f}^{2}+m_{%
\hat{f}}^{2}-m_{\pi }^{2}-m_{\hat{\pi}}^{2}+\hat{\mu}_{1}^{2}-\hat{\mu}%
_{1/2}^{2})\sin \gamma \cos \delta ^{\prime }\sin ^{2}\theta _{1},$ \\ \hline
$g_{f(980)\hat{K}^{+}\hat{K}^{-}}$ & $\frac{1}{\sqrt{2}(a+b)}\left(
m_{f}^{2}+m_{\hat{f}}^{2}-m_{K}^{2}-m_{\hat{K}}^{2}\right) \left( \sin
\gamma +\sqrt{2}\cos \gamma \right) \cos \delta ^{\prime }\sin ^{2}\theta
_{1/2},$ \\ \hline
$g_{f(980)\hat{\kappa}^{+}\hat{\kappa}^{-}}$ & $-\frac{1}{\sqrt{2}(b-a)}%
\left( m_{f}^{2}+m_{\hat{f}}^{2}-m_{\kappa }^{2}-m_{\hat{\kappa}}^{2}\right)
\left( \sin \gamma -\sqrt{2}\cos \gamma \right) \cos \delta ^{\prime }\sin
^{2}\phi _{1/2},$ \\ \hline
$g_{f(1370)\pi ^{+}\pi ^{-}}$ & $\frac{1}{\sqrt{2}a}(m_{\sigma }^{2}+m_{\hat{%
\sigma}}^{2}-m_{\pi }^{2}-m_{\hat{\pi}}^{2}+\hat{\mu}_{1}^{2}-\hat{\mu}%
_{1/2}^{2})\cos \gamma \sin \delta \cos ^{2}\theta _{1},$ \\ \hline
$g_{f(1370)K^{+}K^{-}}$ & $\frac{1}{\sqrt{2}(a+b)}\left( m_{\sigma }^{2}+m_{%
\hat{\sigma}}^{2}-m_{K}^{2}-m_{\hat{K}}^{2}\right) \left( \cos \gamma -\sqrt{%
2}\sin \gamma \right) \sin \delta \cos ^{2}\theta _{1/2},$ \\ \hline
$g_{f(1370)\kappa ^{+}\kappa ^{-}}$ & $-\frac{1}{\sqrt{2}(b-a)}\left(
m_{\sigma }^{2}+m_{\hat{\sigma}}^{2}-m_{\kappa }^{2}-m_{\hat{\kappa}%
}^{2}\right) \left( \cos \gamma +\sqrt{2}\sin \gamma \right) \sin \delta
\cos ^{2}\phi _{1/2},$ \\ \hline
$g_{f(1370)\hat{\pi}^{+}\hat{\pi}^{-}}$ & $\frac{1}{\sqrt{2}a}(m_{\sigma
}^{2}+m_{\hat{\sigma}}^{2}-m_{\pi }^{2}-m_{\hat{\pi}}^{2}+\hat{\mu}_{1}^{2}-%
\hat{\mu}_{1/2}^{2})\cos \gamma \sin \delta \sin ^{2}\theta _{1},$ \\ \hline
$g_{f(1370)\hat{K}^{+}\hat{K}^{-}}$ & $\frac{1}{\sqrt{2}(a+b)}\left(
m_{\sigma }^{2}+m_{\hat{\sigma}}^{2}-m_{K}^{2}-m_{\hat{K}}^{2}\right) \left(
\cos \gamma -\sqrt{2}\sin \gamma \right) \sin \delta \sin ^{2}\theta _{1/2},$
\\ \hline
$g_{f(1370)\hat{\kappa}^{+}\hat{\kappa}^{-}}$ & $-\frac{1}{\sqrt{2}(b-a)}%
\left( m_{\sigma }^{2}+m_{\hat{\sigma}}^{2}-m_{\kappa }^{2}-m_{\hat{\kappa}%
}^{2}\right) \left( \cos \gamma +\sqrt{2}\sin \gamma \right) \sin \delta
\sin ^{2}\phi _{1/2},$ \\ \hline
$g_{f(1500)\pi ^{+}\pi ^{-}}$ & $\frac{1}{\sqrt{2}a}(m_{f}^{2}+m_{\hat{f}%
}^{2}-m_{\pi }^{2}-m_{\hat{\pi}}^{2}+\hat{\mu}_{1}^{2}-\hat{\mu}%
_{1/2}^{2})\sin \gamma \sin \delta ^{\prime }\cos ^{2}\theta _{1},$ \\ \hline
$g_{f(1500)K^{+}K^{-}}$ & $\frac{1}{\sqrt{2}(a+b)}\left( m_{f}^{2}+m_{\hat{f}%
}^{2}-m_{K}^{2}-m_{\hat{K}}^{2}\right) \left( \sin \gamma +\sqrt{2}\cos
\gamma \right) \sin \delta ^{\prime }\cos ^{2}\theta _{1/2},$ \\ \hline
$g_{f(1500)\kappa ^{+}\kappa ^{-}}$ & $-\frac{1}{\sqrt{2}(b-a)}\left(
m_{f}^{2}+m_{\hat{f}}^{2}-m_{\kappa }^{2}-m_{\hat{\kappa}}^{2}\right) \left(
\sin \gamma -\sqrt{2}\cos \gamma \right) \sin \delta ^{\prime }\cos ^{2}\phi
_{1/2},$ \\ \hline
$g_{f(1500)\hat{\pi}^{+}\hat{\pi}^{-}}$ & $\frac{1}{\sqrt{2}a}(m_{f}^{2}+m_{%
\hat{f}}^{2}-m_{\pi }^{2}-m_{\hat{\pi}}^{2}+\hat{\mu}_{1}^{2}-\hat{\mu}%
_{1/2}^{2})\sin \gamma \sin \delta ^{\prime }\sin ^{2}\theta _{1},$ \\ \hline
$g_{f(1500)\hat{K}^{+}\hat{K}^{-}}$ & $\frac{1}{\sqrt{2}(a+b)}\left(
m_{f}^{2}+m_{\hat{f}}^{2}-m_{K}^{2}-m_{\hat{K}}^{2}\right) \left( \sin
\gamma +\sqrt{2}\cos \gamma \right) \sin \delta ^{\prime }\sin ^{2}\theta
_{1/2},$ \\ \hline
$g_{f(1500)\hat{\kappa}^{+}\hat{\kappa}^{-}}$ & $-\frac{1}{\sqrt{2}(b-a)}%
\left( m_{f}^{2}+m_{\hat{f}}^{2}-m_{\kappa }^{2}-m_{\hat{\kappa}}^{2}\right)
\left( \sin \gamma -\sqrt{2}\cos \gamma \right) \sin \delta ^{\prime }\sin
^{2}\phi _{1/2},$ \\ \hline
\end{tabular}%
\end{equation*}%
Again, couplings for light scalars to light pseudoscalars listed in Table IV
reduce to the linear sigma model ones when we decouple heavy fields. In this
sense, we would like to notice that the naive mixing factor used in \cite{LN}
for $g_{f(980)\kappa ^{+}\kappa ^{-}}$ contains an incorrect minus sign 
\footnote{%
In \cite{LN} the mixing factor was assumed to be $\sin \varphi _{s}+\sqrt{2}%
\cos \varphi _{s},$ where $\varphi _{s}$ denotes the scalar mixing angle in
the strange-non-strange basis. A complete calculation yields actually $\sin
\varphi _{s}-\sqrt{2}\cos \varphi _{s}$ which coincides with the present
calculation when we decouple heavy scalars. The angle $\varphi _{s}$
coincides with the angle $\gamma $ in the present work.}. Eventhough
contributions coming from charged $\kappa $ loops are suppressed, they are
crucial in the understanding of two photon decay of the $f(980)$. Indeed,
the sign assumed in \cite{LN} takes the result arising from kaon and pion
loops in the right direction to match experiment. We recalculated this decay
in the L$\sigma $M correcting for this sign finding a $\kappa $ contribution
in the opposite direction. The different contributions to this form factor
as obtained in the L$\sigma $M are quoted in the Table VI. Predictions of
the chiral model for $\overline{q}q$ and $\overline{q}\overline{q}qq$ states
for the different contributions to the form factor describing $%
f_{0}(980)\rightarrow \gamma \gamma $ decay are also shown in the table V.

\begin{center}
Table V
\end{center}
\begin{center}

\begin{tabular}{|l|l|l|l|l|}
\hline
{\small Contr.} & $\left\vert V^{f(980)}\right\vert $ & $\left\vert
V^{f(1500)}\right\vert $ & $\left\vert V^{f(600)}\right\vert $ & $\left\vert
V^{f(1370)}\right\vert $ \\ \hline
${\small K}$ & ${\small 0.409\pm 0.127}$ & ${\small 0.283\pm 0.094}$ & $%
{\small 0.014\pm 0.020}$ & ${\small 0.011\pm 0.016}$ \\ \hline
${\small K,\pi }$ & ${\small 0.601\pm 0.133}$ & ${\small 0.254\pm 0.085}$ & $%
{\small 1.487\pm 0.928}$ & ${\small 0.060\pm 0.069}$ \\ \hline
${\small K,\pi ,\kappa }$ & ${\small 0.662\pm 0.145}$ & ${\small 0.306\pm
0.134}$ & ${\small 1.445\pm 0.937}$ & ${\small 0.102\pm 0.103}$ \\ \hline
${\small K,\kappa ,\pi ,\hat{K}}$ & ${\small 0.664\pm 0.144}$ & ${\small %
0.308\pm 0.135}$ & ${\small 1.444\pm 0.937}$ & ${\small 0.102\pm 0.104}$ \\ 
\hline
${\small K,\kappa ,\pi ,\hat{K},\hat{\pi}}$ & ${\small 0.664\pm 0.144}$ & $%
{\small 0.308\pm 0.135}$ & ${\small 1.444\pm 0.937}$ & ${\small 0.102\pm
0.104}$ \\ \hline
{\small Total} & ${\small 0.682\pm 0.149}$ & ${\small 0.323\pm 0.152}$ & $%
{\small 1.431\pm 0.939}$ & ${\small 0.116\pm 0.107}$ \\ \hline
\end{tabular}
\end{center}

whereas in the linear sigma model \cite{LN} we obtain results listed in
Table VI

\begin{center}
Table VI 
\end{center}
\begin{center}
\begin{tabular}{|l|l|l|l|}
\hline
{\small Contr.} & $K$ & $K,\pi $ & $K,\pi ,\kappa $ \\ \hline
$\left\vert V^{f(980)}\right\vert _{L\sigma M}$ & ${\small 0.446\pm 0.115}$
& ${\small 0.779\pm 0.200}$ & ${\small 0.898\pm 0.135}$ \\ \hline
\end{tabular}
\end{center}

The form factors in Table V yield the following decay widths 
\begin{align}
\Gamma (\sigma & \rightarrow \gamma \gamma )=0.470\pm 0.665\text{ KeV} \\
\Gamma (f(1370)& \rightarrow \gamma \gamma )=0.071\pm 0.155\text{ KeV} \\
\Gamma (f(1500)& \rightarrow \gamma \gamma )=0.741\pm 0.778\text{ KeV.}
\end{align}

\section{Conclusions}

In this paper we work out the predictions of the chiral model for $\overline{%
q}q$ and $\overline{q}\overline{q}qq$ states \cite{NR,NR1} for the two
photon decay of all neutral scalars below 1.5 GeV. Except for the $%
a_{0}(980) $ and the $f_{0}(980)$ , there is no experimental information on
these decays. The predictions of the model for the $a_{0}(980)\rightarrow
\gamma \gamma $ decay are in very good agreement with experiment. As for the 
$f_{0}(980)\rightarrow \gamma \gamma $ we recalculate and update the linear
sigma model predictions for this decay finding a discrepancy with the
experimental results. The situation is improved by the mixing inherent to
the chiral model for $\overline{q}q$ and $\overline{q}\overline{q}qq$ states 
\cite{NR,NR1}. Nevertheless, on the one side the extraction of the width
from experimental data is not an easy task as shown in \cite{Pennington},
thus the world average quoted in \cite{PDG04} should be taken with some
care; on the other side we expect the $f_{0}(980)$ to arise actually as a
mixing of $\overline{q}q$, $\overline{q}\overline{q}qq$ and glueball. The
latter has not been included in the model under consideration and according
to recent analysis based on chiral symmetry the $f_{0}(980)$ can acquire
some component along the glueball direction \cite{Fariborz,Teshima}, thus we
expect modifications to the present picture upon the inclusion of glueball
degrees of freedom. Finally the model gives definite predictions for the two
photon decays of the $f_{0}(600),$ $f_{0}(1370)$ and $f_{0}(1500) $. It is
particularly interesting the small decay width of the latter two mesons into
two photons, even if they are composed of quarks in this model. The small
width of these mesons has been usually argued as the signal for a large
glueball component.

\section{Acnowledgments}

Work supported by Conacyt- M\'{e}xico under project 37234-E.

\end{document}